# Photonic Thermal Conduction by Infrared Plasmonic Resonators in Semiconductor Nanowires

Eric J. Tervo,[1,a)] Michael E. Gustafson,[2] Zhuomin M. Zhang,[1] Baratunde A. Cola,[1,3] and Michael A. Filler[2,b)]

[1]George W. Woodruff School of Mechanical Engineering, Georgia Institute of Technology, Atlanta, Georgia 30332, USA

[2]School of Chemical & Biomolecular Engineering, Georgia Institute of Technology, Atlanta, Georgia 30332, USA

[3]School of Materials Science and Engineering, Georgia Institute of Technology, Atlanta, Georgia 30332, USA

Photons typically do not contribute to thermal transport within a solid due to their low energy density and tendency to be quickly absorbed. We propose a practical material system – infrared plasmonic resonators embedded in a semiconductor nanowire – that leverages near-field electromagnetic coupling to achieve photonic thermal transport comparable to the electronic and phononic contributions. We analytically show photonic thermal conductivities up to about 1 W m$^{-1}$ K$^{-1}$ for 10 nm diameter Si and InAs nanowires containing repeated resonators at 500 K. The nanowire system outperforms plasmonic particles in isotropic environments and presents a pathway for photonic thermal transport to exceed that of phonons and electrons.

Thermal transport in a nanoscale material or composite is typically dominated by electronic contributions in metals and phononic contributions in insulators.[1] Photonic effects, on the other hand, are rarely considered for thermal transport within a material due to their low far-field energy density and tendency towards fast absorption in most media.[2] When bodies are separated by nanoscale gaps, however, the more energy dense electromagnetic near-fields interact and enable thermal radiation exceeding the far-field limit.[3,4] Near-field effects have received considerable attention in the past decade for their applications in thermal management,[5] heat-assisted magnetic recording,[6] and energy conversion,[7] among others.[8] Most of this research has focused on thermal radiation between two bodies across a single gap, and not the bulk thermal conductivity of the material or composite itself.

An alternative approach is to utilize near-field effects between many nanostructures, such as in nanofluids,[9] packed nanoparticle beds,[10] or ordered nanostructure arrays.[11] Coupling of the electromagnetic fields between neighboring polaritonic particles can lead to energy transport along a particle chain or in an array.[9-26] This effect has been explored for both sub-diffractional waveguiding[12-20] and thermal conduction by the plasmonic modes.[9-11,21-25] Previous research on thermal conduction by coupled polaritonic modes has several limitations. One is the assumption of ballistic propagation,[11,25] which is only applicable to repeated particles of nanoscale array lengths due to typical propagation lengths on the order of tens to hundreds of nanometers. Another limitation is choosing optimized particle shapes and orientations,[11,24,25] which may be difficult to fabricate or assemble. Prior research has solely explored particles immersed in homogeneous environments. Even for optimized geometries, predicted maximum thermal conductivities due to the photonic component are only about 0.1 W m$^{-1}$ K$^{-1}$ for spherical, physically-contacting copper particles 10 nm in diameter at 900 K[21] and about 0.05 W m$^{-1}$ K$^{-1}$ for prolate spheroidal,

---

a)Email: eric.tervo@gatech.edu
b)Email: mfiller@gatech.edu

physically-contacting SiC particles 50 nm in minor axis diameter at 700 K.[24] An experimental study of packed nanoparticles found significantly enhanced thermal conduction due to polaritonic coupling, but existing models could not fully explain these results.[10] A nanostructured material system has not yet been designed that promises significant radiation contributions to thermal transport.

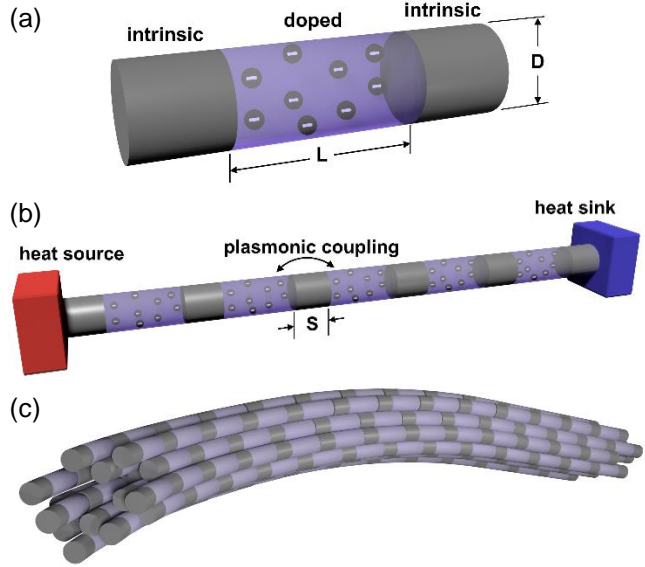

FIG. 1. Schematics of (a) a n-type doped plasmonic resonator embedded in an intrinsic semiconductor nanowire, (b) photonic thermal conduction by coupled plasmonic resonators, and (c) a hypothetical nanowire bundle which could transport heat by thermal radiation.

In this letter, we propose and analyze a structure where the photonic contributions to the thermal conductivity are comparable to electronic and phononic conduction: doped semiconductor plasmonic resonators embedded in an intrinsic semiconductor nanowire. This structure is shown schematically in Figures 1(a) and 1(b). We consider a nanowire of diameter $D$ containing many resonators such that the photonic thermal transport is diffusive. The resonators have length $L$ and are separated by distance $S$ or center-to-center spacing $d = L + S$. The plasmonic resonators confine thermal infrared light to deep subwavelength volumes, providing high photonic energy density,[27] and the anisotropic nanowire environment focuses and enhances the fields between resonators.[28] In our previous work, we showed that this nanowire structure leads to stronger coupling, longer propagation lengths, and higher group velocities compared to the same resonator chain in isotropic surroundings.[20] These structures may be grown very precisely[29,30] with the bottom-up vapor-liquid-solid method,[31] which could facilitate low-cost production of macroscale composites containing many nanowires, as depicted in Figure 1(c). Doped plasmonic resonators also offer the potential for tunability via optical pumping or electrostatic gating of the adjacent intrinsic regions,[32,33] which could lead to actively-modulated thermal transport. Here, we provide details of the materials being analyzed and the methods used to calculate photonic thermal conductivity with a kinetic theory formalism. We then present results on the effects of nanowire geometry in order to optimize the photonic component, showing that small diameter nanowires with long resonators can achieve photonic thermal conductivities up to about 1 W m$^{-1}$ K$^{-1}$ at 500 K. We conclude with a discussion of possible future work and the potential of this type of composite material.

Kinetic theory is appropriate to describe radiative thermal transport when modes propagate at least between two adjacent particles.[26,34-36] The dispersion for propagating plasmons along the chain must first be obtained, and the absorption spectra method we previously developed[20] provides this information from numerical electromagnetic scattering simulations or absorption spectroscopy measurements. We use this method because it is valid for plasmonic resonators in arbitrary nonhomogeneous environments such as nanowires, and it allows the dispersion of a long chain to be calculated from absorption characteristics of just two resonators. We use the discrete dipole approximation (DDA) with the DDSCAT software[37] using the FLTRCD method to obtain infrared absorption efficiency, defined as the absorption cross section divided by the cross-sectional area of an equivalent volume sphere, for a single resonator as well as two resonators embedded in a nanowire with varying $S$. Representative absorption efficiencies for InAs nanowires are shown in Figure 2(a). The illumination is longitudinally polarized along the nanowire axis, because the nanowire geometry suppresses transverse modes and causes longitudinal modes to remain dipolar even at very small $S$.[28,29,38] The materials and optical properties are taken as either intrinsic Si[39] with n-type doped $3\times10^{20}$ cm$^{-3}$ Si resonators[40] or intrinsic InAs[39] with n-type doped $3.25\times10^{19}$ cm$^{-3}$ InAs resonators,[41] which are experimentally achievable. It is important to verify the applicability of optical data or models over the entire frequency range to be examined, and we have done so for these materials. Si is an attractive material for its low cost and scalability, while InAs provides an interesting comparison because of its low loss and good carrier mobility, which has motivated studies of InAs for plasmonic devices.[42] After the absorption peak positions are determined for several separations, the data are fit to the so-called plasmon ruler equation[43-46]

$$\frac{\omega_1 - \omega_2}{\omega_1} = A_0 \exp\left(\frac{-S/L}{\tau}\right) \quad (1)$$

where $\omega_1$ is the absorption peak angular frequency for a single resonator and $\omega_2$ is the absorption peak angular frequency for two resonators with a spacing $S$ and length $L$. $A_0$ is a dimensionless proportionality constant, and $\tau$ is the decay length scaling factor, which is a measure of the attenuation distance of the resonator's electric field. This fitting process is depicted for one case in Figure 2(b). With $A_0$ and $\tau$, $\omega_2$ may be extrapolated for any desired separation distance. This, in turn, allows the calculation of the coupling strength between two resonators for any separation distance:[20]

$$\omega_c^2 = \omega_1^2 - \omega_2^2 \quad (2)$$

The dispersion may finally be determined for a resonator chain with $N$ neighboring resonators on either side of the central one with the equation[20]

$$\omega^2 = \omega_1^2 - \frac{\xi}{4} - 2\sum_{n=1}^{N} \omega_{c,n} \cos(nkd) \quad (3)$$

where $\omega$ is the angular frequency, $\xi$ is the full-width at half maximum of the single resonator absorption peak, $\omega_{c,n}$ is the coupling strength between the center resonator and its $n$th nearest neighbor, and $k$ is the wavevector varied from zero to the first Brillouin zone $\pi/d$. All calculations are performed for $N = 15$ nearest neighbors, which is sufficient for the calculations to converge. The dispersion for one representative case is shown in Figure 2(c).

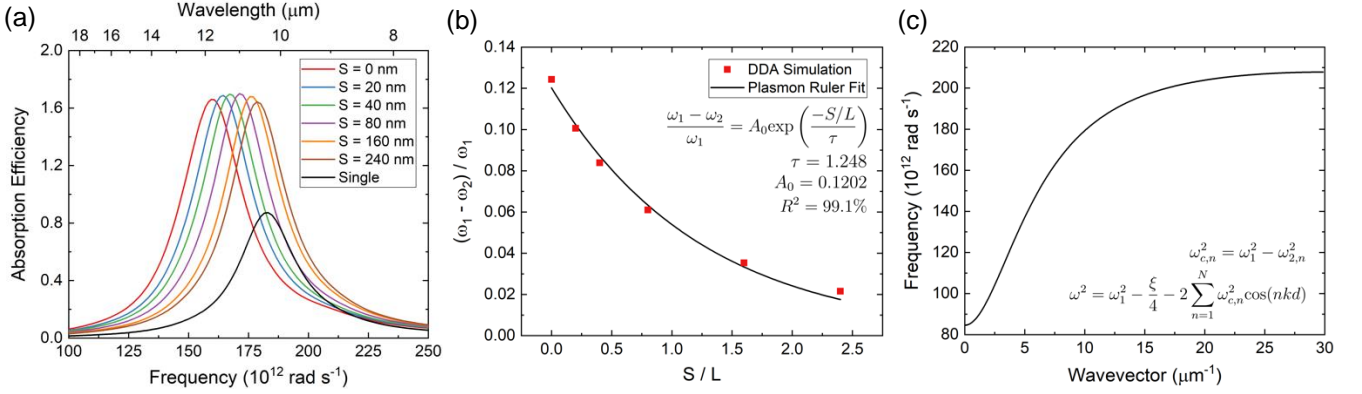

FIG. 2. (a) Absorption spectra from DDA calculations, (b) absorption peaks fit to the plasmon ruler equation (equation inset), and (c) dispersion relation (dispersion and coupling strength equations inset) for InAs resonators in an intrinsic InAs nanowire. All resonators have $L = 100$ nm in a wire of $D = 100$ nm, and the dispersion is for a separation distance of $S = 5$ nm.

The dispersion provides the inputs needed to calculate thermal conductivity, namely the group velocity $v_g = \partial \omega / \partial k$ and the propagation length $\Lambda = v_g (4\xi)^{-1}$. Kinetic theory gives the following equation for the thermal conductivity[24]

$$\kappa = \frac{1}{\pi A} \int \hbar \omega \Lambda \frac{\partial f_{BE}}{\partial T} d\omega \qquad (4)$$

where $A = \pi D^2/4$ is the cross-sectional area of the nanowire, $f_{BE}$ is the Bose-Einstein distribution, and $T$ is the temperature. All calculations are performed for $T = 500$ K. This equation assumes that all the resonators are near thermal equilibrium, so the temperature gradient is small. Large temperature differences along the nanowire could lead to effects such as nonlinear temperature gradients, and the effective thermal conductivity would have to be calculated with a different method.[22,47]

We first evaluate the effect of the nanowire environment by comparing the nanowire system shown in Figure 1(b) to the same chain of plasmonic resonators immersed in an isotropic vacuum. For this latter case, the geometry and spacing of the resonators is unchanged, but the intrinsic regions illustrated in Figure 1(b) are replaced by vacuum such that the resonators are 'floating' in space. These calculations are performed for both Si and InAs, and the results are shown in Figure 3 for $D = 100$ nm and $S = 5$ nm. These dimensions are used as a starting point because nanowires of this diameter have been grown previously[28,29] and small separation distances exhibit good electromagnetic coupling.[20] For both materials, the spectral thermal conductivity in Figure 3(a) and 3(b) is an order of magnitude higher when the resonators are embedded in the nanowire, resulting in total thermal conductivities of $1.1 \times 10^{-3}$ ($1.56 \times 10^{-4}$) W m$^{-1}$ K$^{-1}$ in vacuum and $1.46 \times 10^{-2}$ ($1.29 \times 10^{-2}$) W m$^{-1}$ K$^{-1}$ in the nanowire for InAs (Si). Because $\xi$ does not change appreciably with the resonator environment,[20] the increase in thermal conductivity is a direct result of stronger coupling. This leads to higher group velocities as seen in Figure 3(c) and 3(d) and longer propagation lengths as shown in Figure 3(e) and 3(f). Propagation length clearly plays a significant role in determining the total thermal conductivity, as illustrated in Equation (4), but the spectral region where the surface plasmons propagate also influences the results through the $\hbar\omega$ and $\partial f_{BE}/\partial T$ factors. Higher frequencies carry more energy, but those states also have a lower thermal population of carriers as given by the Bose-Einstein distribution. We plot the combined $\hbar\omega \, \partial f_{BE}/\partial T$ in Figure S1, which shows that from a thermal transport perspective it is always preferable to have lower frequency modes if this does not adversely affect the coupling strength.

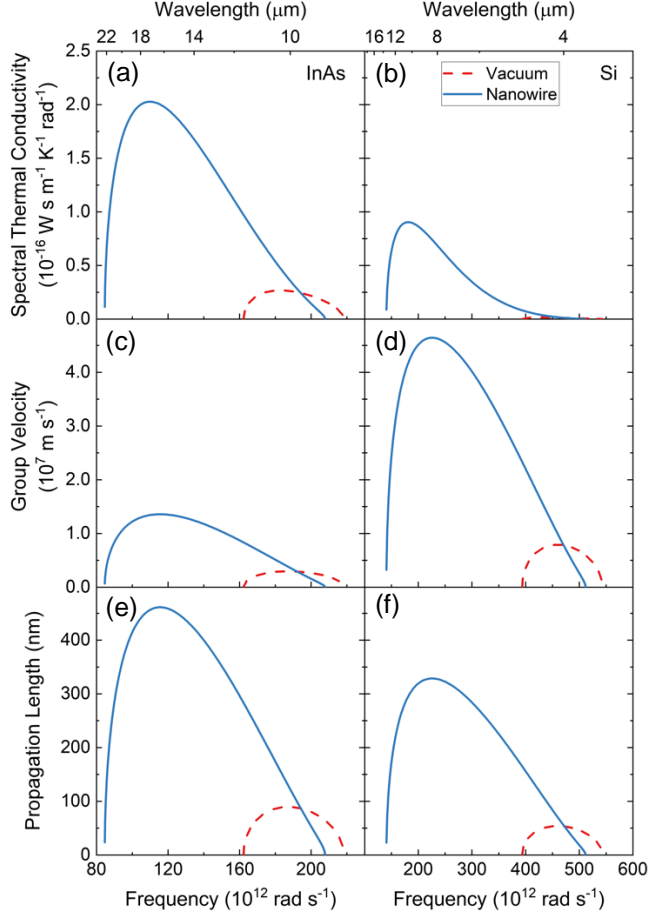

FIG. 3. (a,b) Spectral thermal conductivity of a chain of plasmonic resonators in a nanowire (solid) and in isotropic vacuum (dashed). Higher thermal conductivity results from enhanced group velocities (c,d) and propagation lengths (e,f) for both InAs (a,c,e) and Si (b,d,f) systems with $D = 100$ nm and $S = 5$ nm.

To optimize photonic thermal transport in these nanowire systems, we investigate the impact of nanowire and resonator geometry by varying $D$ and $L$. Figure 4(a) shows the total photonic thermal conductivity for InAs and Si versus nanowire diameter for $S = 5$ nm, either fixing $L = 100$ nm or $AR = LD^{-1} = 1$. We examine diameters down to 10 nm, because $D = 10$ nm has been experimentally demonstrated[48] and smaller $D$ requires more computational resources because of the finer discretization needed with the DDA simulations. We keep $S$ at 5 nm because this is an attainable separation distance for the resonators,[29,30] and because smaller $S$ will lead to stronger coupling[20] and therefore higher photonic thermal conductivity (for example, see Figure S2). At smaller separation distances, the radiation formalism employed may break down as transport transitions from thermal radiation to conduction.[49-51] When $AR$ is constant, we observe a 3.8 times increase in $\kappa$ for InAs and no increase for Si as $D$ decreases from 100 to 10 nm. These results are due to competing effects. $\kappa$ is proportional to $D^{-2}$, which tends to increase $\kappa$ with decreasing $D$. The plasmon ruler equation parameters $\tau$, $A_0$, and $\omega_1$ are not strongly affected, as illustrated in Figures S3 – S6. However, the constant $AR$ means that $L$ decreases with smaller $D$. Decreasing $L$ hampers coupling between resonators because the coupling strength decreases with the exponential of $-S/L$ as shown by Equations (1) and (2). This reduces the propagation lengths for both materials, as illustrated in Figures S7(a) and S8(a), which partially counteracts the $\kappa \propto D^{-2}$ effect for InAs and completely counteracts this effect for Si, as shown in Figures 4, S7(b), and S8(b). The coupling strength and propagation lengths are reduced more for Si than InAs because Si has a higher resonance frequency, which requires a larger change in coupling strength for the same change in $L$ as shown by Equation (1). Doping InAs to the level of that for Si would increase its resonance frequency and the trend in thermal conductivity vs. diameter for constant $AR$ would be closer to the trend for Si.

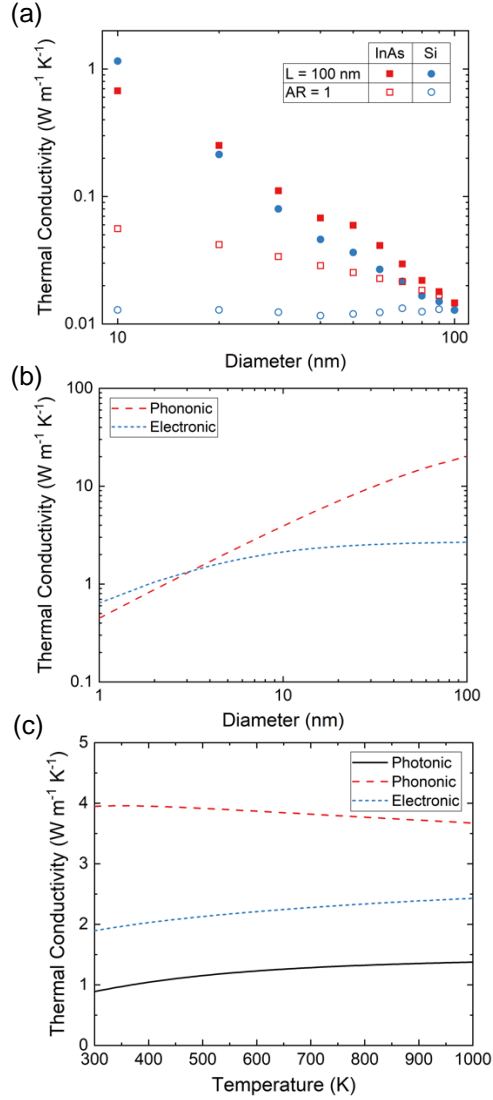

FIG. 4. (a) Photonic thermal conductivity of InAs and Si systems as a function of nanowire diameter for either fixed resonator length or fixed resonator aspect ratio, with $S = 5$ nm in all cases. (b,c) Estimate of phonon and electron thermal conductivity of a single-crystal n-type silicon nanowire with phosphorus concentration $3\times10^{20}$ cm$^{-3}$ for (b) varying diameter at 500 K and (c) varying temperature with 10 nm diameter.

When $L$ is constant, we observe significant increases in $\kappa$ with decreasing $D$ for both InAs and Si, resulting in the largest calculated $\kappa = 1.15$ W m$^{-1}$ K$^{-1}$ for Si at $D = 10$ nm. The increase in $\kappa$ is largely driven by its proportionality to $D^{-2}$, but other factors must also be considered. In the plasmon ruler equation, $A_0$ increases, $\tau$ substantially decreases, and $\omega_1$ decreases as the diameter is reduced with nearly identical trends for InAs and Si, shown by Figures S3 through S6. Decreasing $\tau$ reduces the coupling strength, which is partially counteracted by the increase in $A_0$, which increases the coupling strength. This results in reduced propagation lengths as shown in Figures S7(a) and S8(a), although the reductions are not as drastic as when $AR$ is held constant. In addition to these impacts on the coupling strength and propagation lengths, the dispersions redshift with reduced diameter as illustrated in Figures S7 and S8. This moves the $\omega\, \partial f_{BE}/\partial T$ term of Equation (4) to higher values as indicated in Figure S1 and increases the thermal conductivity. Slightly different enhancements between InAs and Si result from this shift, which can be understood by comparing Figure S1 with Figures S7 and S8. As the InAs dispersion shifts, it moves from a steep portion of the $\omega\, \partial f_{BE}/\partial T$ curve to a shallower portion at lower frequencies. Si, on the other hand, begins at a higher frequency and the majority of the dispersion remains on a steep portion of the $\omega\, \partial f_{BE}\partial T$ curve as the dispersion redshifts. This difference is responsible for the higher thermal conductivity of the Si nanowire at 10 nm diameter, indicating the potential of this common semiconductor.

The photonic thermal conductivity is comparable to that expected for the electronic and phononic contributions in small nanowires. For comparison, we plot in Figure 4(b) and (c) the estimated phonon and electron thermal conductivity of a single-crystal n-type silicon nanowire with phosphorus concentration of $3\times10^{20}$ cm$^{-3}$. We calculate bulk phonon and electron conductivities with the Boltzmann transport equation in the relaxation-time approximation[52-54] using input data from Ohishi *et al.*,[55] which provides good agreement with experimental data for high dopant concentrations. Effective carrier mean free paths were calculated following the method of Dames and Chen,[56] which are used to calculate thermal conductivity of the nanowire with a Boltzmann transport equation approach[57] assuming diffuse boundary scattering.[1] The photonic and phononic/electronic thermal conductivities have opposite trends with diameter, which suggests that small diameter nanowires may be an ideal platform to experimentally resolve photonic contributions to thermal transport. The decreasing phononic and electronic contributions are due to a decreasing effective mean free path as the diameter is reduced. At diameters less than 10 nm, the photonic component may even exceed the phononic and electronic components. The photonic and phononic contributions also have opposite trends with temperature, which suggests that elevated temperature or temperature-dependent tests may permit the measurement of photonic thermal conduction. The phonon contribution decreases due to the increased phonon scattering rate, while the photon part increases according to the thermal population of states as illustrated in Figure S1. The alternating doped/intrinsic regions of our nanowires may also introduce additional scattering events for phonons and electrons, or the structure could be engineered with defects or geometric features[58] to further decrease the phonon and electron thermal conductivities.

Our results demonstrate a pathway towards significant photonic thermal transport in nanoengineered structures, but additional work is needed to identify ideal materials, geometries, and operating conditions for this transport mechanism. Future analytical work should focus on further optimization of coupling between resonators, including investigating the impact of dopant concentration[59] and nonuniform/noncircular nanowire cross-sections, to increase the photonic thermal transport beyond that predicted in this work. The effects of plasmon confinement should also be considered in small diameter structures. Future experimental efforts such as thermal conductivity measurements of single nanowires[60] could confirm our predictions. The possibility of dynamically tuning the thermal transport with gating or illumination[32,33] is another interesting research direction. Efforts focused on large-scale production, alignment, and characterization of the nanowire structures shown in Figure 1(c) could eventually lead to technologically useful macroscale composites.

In summary, we have proposed and theoretically analyzed an experimentally practical nanoscale material structure capable of achieving substantial photonic thermal conduction. Repeated plasmonic resonators in a nanowire lead to electromagnetic coupling and energy transport along the chain. Using DDA simulations and our previously reported absorption spectra method to obtain the surface plasmon dispersion relations, we find photonic thermal conductivities can reach about 1 W m$^{-1}$ K$^{-1}$ for 10 nm diameter nanowires at 500 K, which is a higher thermal conductivity at a lower temperature than previous predictions for particle chains. Because the phononic and electronic contributions to thermal transport decrease with decreasing nanowire diameter and increasing temperature, the photonic effects are non-negligible in this regime and present an opportunity to experimentally observe photonic conduction by surface plasmons. We expect this work to lead to new types of nanoscale materials leveraging photonic effects for thermal transport.

See supplementary material for additional figures showing frequency and temperature behavior of thermal conductivity, the impact of resonator spacing, and how nanowire diameter affects parameters in the plasmon ruler equation as well as mode propagation length.

E.J.T acknowledges support from the National Science Foundation Graduate Research Fellowship Program under Grant No. DGE-1650044, and M.A.F. acknowledges support from the National Science Foundation under Grant No. CBET-1510934. Any opinions, findings, and conclusions or recommendations expressed in this material are those of the authors and do not necessarily reflect the views of the National Science Foundation. Z.M.Z. would like to thank support from the U.S. Department of Energy, Office of Science, Basic Energy Sciences (DE-SC0018369).

# Supplementary Material
Photonic Thermal Conduction by Infrared Plasmonic Resonators in Semiconductor Nanowires

Eric J. Tervo, Michael E. Gustafson, Zhuomin M. Zhang, Baratunde A. Cola, and Michael A. Filler

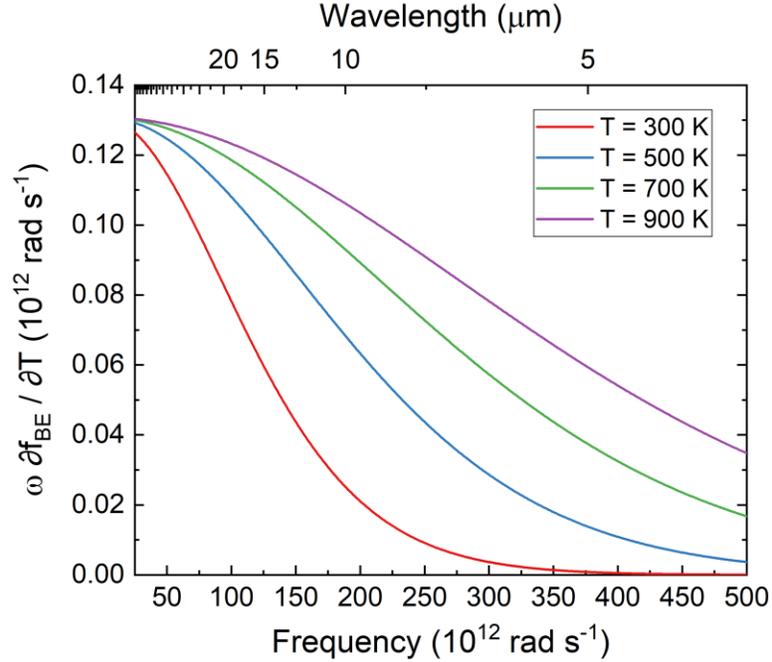

FIG. S1. Contribution of $\omega \, \partial f_{BE}/\partial T$ factors to thermal conductivity (Equation 4) for varying frequency and temperatures. Lower energy propagating modes are preferable for thermal transport.

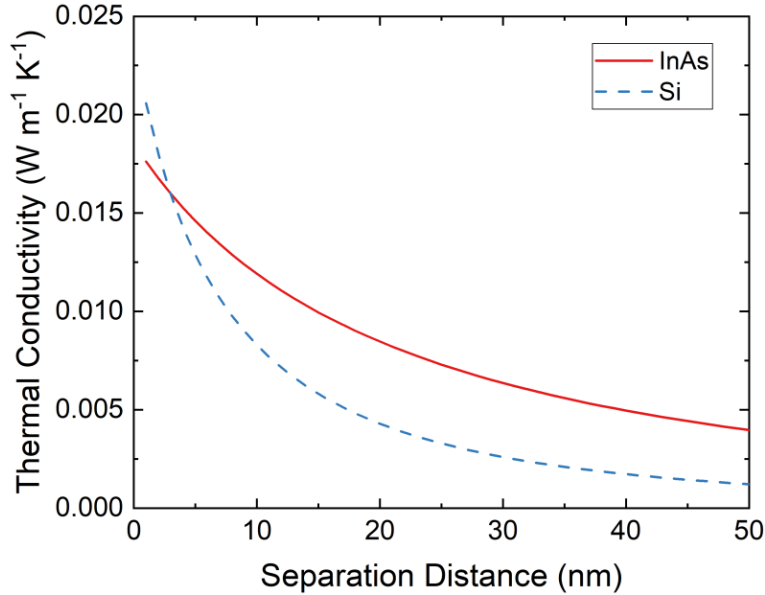

FIG. S2. Impact of separation distance on photonic thermal conductivity for InAs and Si nanowires with $D = 100$ nm and $L = 100$ nm. Although we show distances down to 1 nm, in this small regime the near-field radiation formalism is questionable, so we perform all other calculations at $S = 5$ nm.

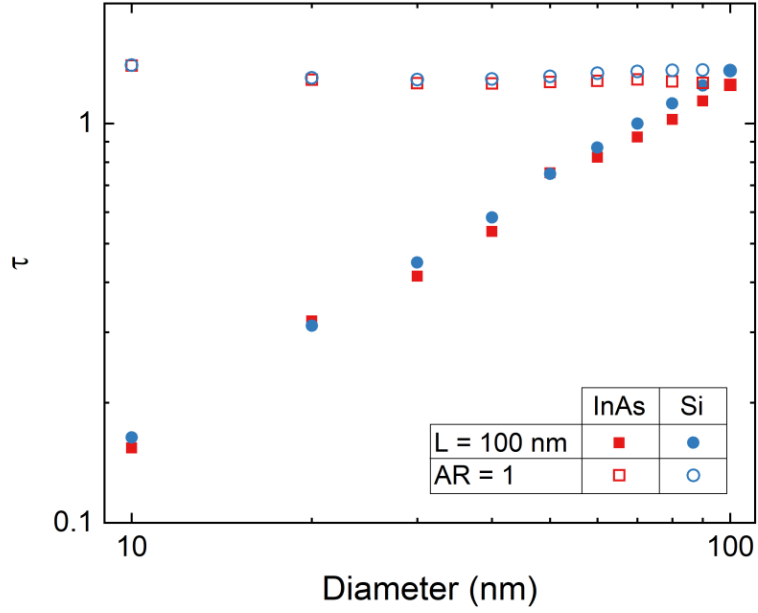

FIG. S3. Decay length scaling factor for InAs and Si nanowires as a function of nanowire diameter for either fixed resonator length or fixed resonator aspect ratio, with $S = 5$ nm in all cases. Diameter has little effect as long as the aspect ratio is constant, but $\tau$ decreases with increasing $AR$ as $D$ decreases.

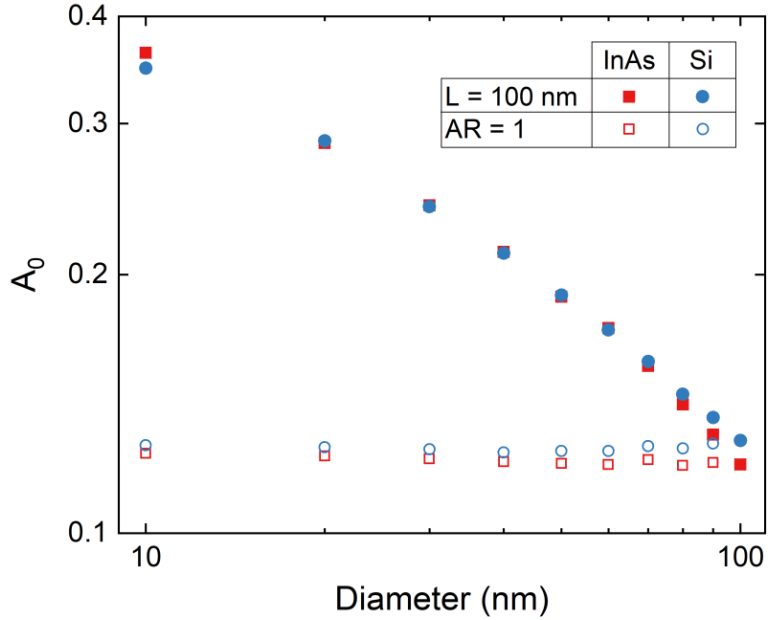

FIG. S4. Proportionality constant in plasmon ruler equation for InAs and Si nanowires as a function of nanowire diameter for either fixed resonator length or fixed resonator aspect ratio, with $S = 5$ nm in all cases. $A_0$ increases substantially as $AR$ increases, which tends to increase photonic conduction.

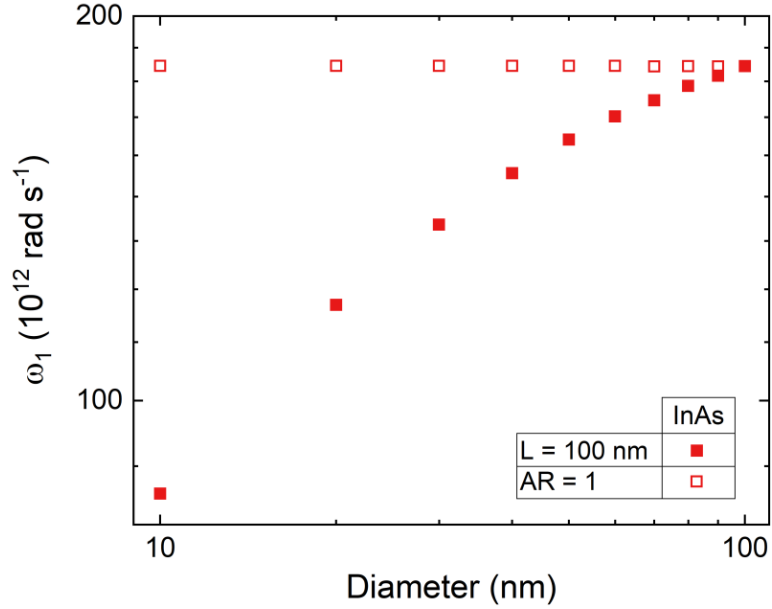

FIG. S5. Single resonator peak absorption frequency for InAs nanowires as a function of nanowire diameter for either fixed resonator length or fixed resonator aspect ratio. Lower frequency resonances with larger *AR* increase thermal transport.

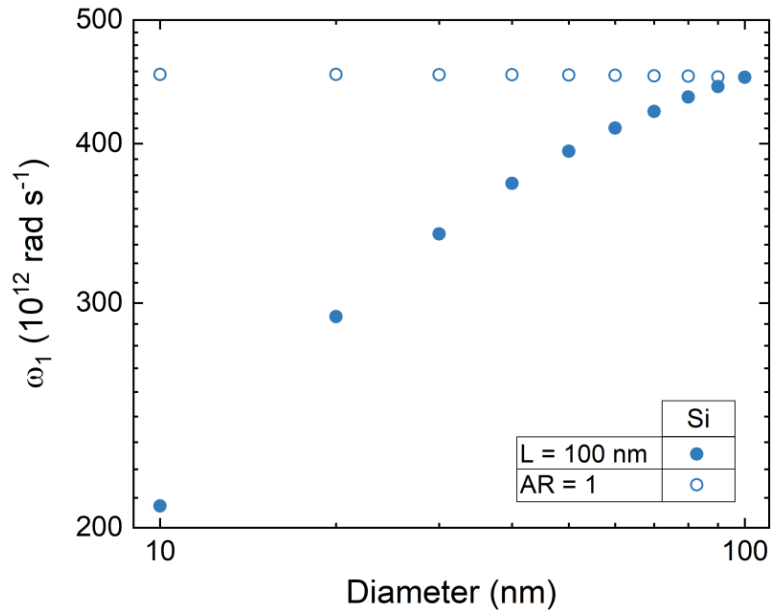

FIG. S6. Single resonator peak absorption frequency for Si nanowires as a function of nanowire diameter for either fixed resonator length or fixed resonator aspect ratio. Lower frequency resonances with larger *AR* increase thermal transport.

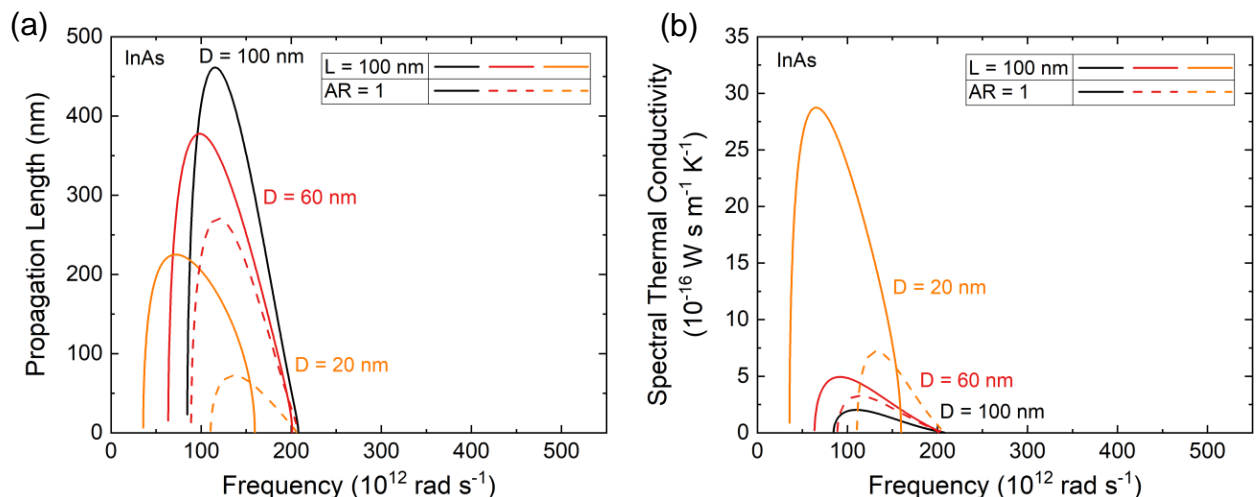

FIG. S7. (a) Propagation lengths and (b) spectral thermal conductivity for InAs nanowires with three different diameters as either $L = 100$ nm (solid lines) or $AR = 1$ (dashed lines) is held constant. The axes share common scales with Figure S8 for comparison. Although propagation length decreases with diameter, the thermal conductivity increases due to its inverse proportionality to nanowire area and the shift of the dispersion to lower frequencies when $L = 100$ nm (see Figure S1).

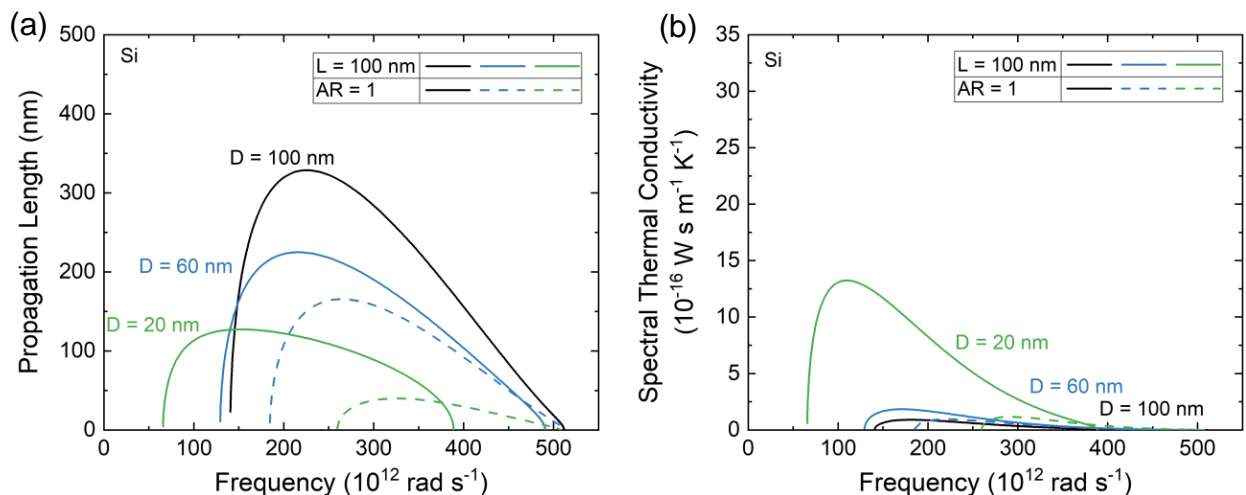

FIG. S8. (a) Propagation lengths and (b) spectral thermal conductivity for Si nanowires with three different diameters as either $L = 100$ nm (solid lines) or $AR = 1$ (dashed lines) is held constant. The axes share common scales with Figure S7 for comparison. Similar effects are seen with InAs nanowires (see Figure S7).

4